\documentclass{article}

\usepackage[preprint]{neurips_2026}

\usepackage[utf8]{inputenc}
\usepackage[T1]{fontenc}
\usepackage{microtype}
\usepackage{graphicx}
\usepackage{subcaption}
\usepackage{booktabs}
\usepackage{hyperref}
\usepackage{url}

\usepackage{amsmath}
\usepackage{amssymb}
\usepackage{mathtools}

\usepackage{xspace}
\usepackage{float}

\usepackage{siunitx}
\DeclareSIUnit{\angstrom}{\text{\smash{\AA}}}
\usepackage{chemformula}
\usepackage{xcolor}
\usepackage{colortbl}
\usepackage{tikz}

\usepackage{acro}

\DeclareAcronym{mlip}{short=MLIP, long=machine learning interatomic potential}
\DeclareAcronym{pes}{short=PES, long=potential energy surface}
\DeclareAcronym{ad}{short=AD, long=automatic differentiation}
\DeclareAcronym{asd}{short=ASD, long=automatic sparse differentiation}
\DeclareAcronym{dft}{short=DFT, long=density functional theory}
\DeclareAcronym{dos}{short=DOS, long=vibrational density of states}
\DeclareAcronym{md}{short=MD, long=molecular dynamics}
\DeclareAcronym{gnn}{short=GNN, long=graph neural network}
\DeclareAcronym{mpnn}{short=MPNN, long=message-passing neural network}
\DeclareAcronym{mof}{short=MOF, long=metal--organic framework}
\DeclareAcronym{cof}{short=COF, long=covalent organic framework}
\DeclareAcronym{jvp}{short=JVP, long=Jacobian-vector product}
\DeclareAcronym{vjp}{short=VJP, long=vector-Jacobian product}
\DeclareAcronym{hvp}{short=HVP, long=Hessian-vector product}

\newcommand{\ad}{\ac{ad}\xspace}
\newcommand{\asd}{\ac{asd}\xspace}
\newcommand{\dft}{\ac{dft}\xspace}

\newcommand{\pes}{\ac{pes}\xspace}

\newcommand{\gnns}{\acp{gnn}\xspace}
\newcommand{\mpnn}{\ac{mpnn}\xspace}
\newcommand{\mpnns}{\acp{mpnn}\xspace}

\newcommand{\mlip}{\ac{mlip}\xspace}
\newcommand{\mlips}{\acp{mlip}\xspace}

\newcommand{\mofs}{\acp{mof}\xspace}

\newcommand{\cofs}{\acp{cof}\xspace}

\newcommand{\jvps}{\acp{jvp}\xspace}
\newcommand{\vjp}{\ac{vjp}\xspace}
\newcommand{\vjps}{\acp{vjp}\xspace}
\newcommand{\hvp}{\ac{hvp}\xspace}
\newcommand{\hvps}{\acp{hvp}\xspace}

\newcommand{\model}[1]{\textsc{#1}\xspace}
\newcommand{\mace}{\model{Mace}}
\newcommand{\pet}{\model{Pet}}

\newcommand{\torch}{PyTorch\xspace}
\newcommand{\jax}{JAX\xspace}

\newcommand{\rscan}{r$^2$SCAN\xspace}

\providecommand{\reals}{\mathbb{R}}
\providecommand{\bigo}[1]{\mathcal{O}(#1)}

\providecommand{\CV}{C_V}

\let\R\undefined
\newcommand{\R}{\boldsymbol{r}}

\newcommand{\Rc}{r}

\let\V\undefined
\newcommand{\V}{\boldsymbol{v}}

\newcommand{\hess}{\mathbf{H}}
\newcommand{\adj}{\mathbf{A}}

\newcommand{\seed}{\mathbf{S}}
\newcommand{\comp}{\hess \seed}

\definecolor{exactshade}{gray}{0.955}

\newcommand{\ablFpMedCeil}{5\times10^{-5}}
\newcommand{\ablProdMedHi}{0.12}

\newcommand{\ablDispStruct}{RSM0274}
\newcommand{\ablDispShift}{67}
\newcommand{\ablDispShiftCv}{1.0}

\usepackage[capitalize,noabbrev]{cleveref}

\title{Truncated automatic sparse differentiation\\for machine learning interatomic potentials}

\author{%
  Marcel~F.~Langer$^{1}$ \quad Adrian~Hill$^{2,3}$ \quad Michele~Ceriotti$^{1}$\\[0.6em]
  $^{1}$Laboratory of Computational Science and Modeling (COSMO), EPFL, Lausanne, Switzerland\\
  $^{2}$Machine Learning Group, Technical University of Berlin, Berlin, Germany\\
  $^{3}$BIFOLD -- Berlin Institute for the Foundations of Learning and Data, Berlin, Germany\\[0.3em]
  \texttt{\{marcel.langer,michele.ceriotti\}@epfl.ch}, \texttt{hill@tu-berlin.de}%
}

\begin{document}

\maketitle

\begin{abstract}
Machine learning interatomic potentials (MLIPs) learn the mapping from atomic positions to potential energy.
The forces, the negative gradient of this energy, drive molecular dynamics and are readily obtained using automatic differentiation.
Higher-order derivatives, most notably the Hessian, describe collective motion and allow the direct prediction of experimental observables, but are considered computationally inaccessible for large systems.
We suggest a solution: in physical systems, interactions decay with distance, and most MLIPs build on this locality through message passing up to a finite receptive field.
This implies both sparsity of higher-order derivatives and their decay with distance.
This structure can be exploited using automatic sparse differentiation (ASD).
We explain how to compute the sparsity pattern for MLIP derivatives and demonstrate that, for multiple foundation MLIPs, ASD computes full Hessians of large porous materials exactly, but with modest speedups at best.
The larger gains come from truncated ASD: discarding small, but nonzero, Hessian entries between distant atoms yields order-of-magnitude speedups with negligible impact on predicted observables.
\end{abstract}

\section{Introduction}
\label{sec:intro}

\Acp{mlip}, i.e., learned approximations to the \citet{bo27} \pes based on first-principles reference data, have become increasingly valuable tools for computational materials science, chemistry, and biology~\citep{uctm21,dcc19}.
In practice, not only the potential energy, but also its derivatives carry essential information for atomistic modeling.
First derivatives describe forces acting on atoms and allow the simulation of atomic motion through molecular dynamics.
Higher-order derivatives describe the curvature of the \pes. Under the assumption of small deviations around a local minimum (the \emph{harmonic approximation}), the Hessian gives rise to collective modes of motion (the \emph{phonons}), a long-established model of vibrations in molecules and solids that underpins the prediction of experimental observables such as heat capacity, vibrational spectra, and thermal transport~\citep{d12,d93}.
Third and higher-order derivatives are \emph{anharmonic} terms, interpreted as scattering between phonon modes and determining their lifetimes and linewidths~\citep{mf62}.

\looseness=-1 Obtaining derivatives of \mlips is therefore an important task. Since the total energy is a scalar, \ad allows the calculation of forces (its negative gradient) in a single backward pass, at the same asymptotic cost as the energy prediction itself. This is not the case for higher-order derivatives: if the forward pass is $\bigo{N}$, as for the \mlips considered here, the full Hessian takes $3N$ derivative passes of $\bigo{N}$ each, hence $\bigo{N^2}$; the full third-derivative tensor $(3N)^2$ passes, hence $\bigo{N^3}$, and so on.
This bottleneck has restricted the use of \mlips for vibrational analysis to modest system sizes thus far~\citep{grm25,ekzp25,lsbm25}.

Fortunately, this asymptotic scaling is a worst-case estimate for \emph{generic} higher-order derivatives. If derivatives have additional structure---in particular, sparsity---they can be computed more efficiently by omitting known zeros. This technique is called \asd~\citep{cpr74,pt79}.
This work gives a blueprint for using \asd with \mlips based on \mpnns. We show how to compute the sparsity pattern for a given \mlip in closed form, and argue that this procedure also gives rise to a hierarchy of truncated $k$-hop sparsity patterns that converges to the exact sparsity pattern of the model. For common \mlips (\mace, \pet) and porous materials---\mofs, \cofs, and zeolites---we show empirically that (a) \asd with the exact pattern recovers dense Hessians to floating-point accuracy at modest speedups, larger for models with smaller receptive fields, (b) Hessian entries decay rapidly with hop distance, at a rate set by the chemistry of the system, and (c) truncating the pattern at a low hop count therefore trades a controlled approximation for much larger speedups, with often negligible impact on predicted observables.
Our results bring experimental observables of large systems within routine reach, opening the door to systematic evaluation of \mlips on, and fine-tuning with, real-world measurement results.

\section{Background}
\label{sec:background}

\paragraph{Machine learning interatomic potentials}
Under the \citet{bo27} approximation, the $N$ nuclei of a molecule or material move on a \pes $E = E(\{(\R_i, Z_i)\}_{i=1}^{N})$ that depends only on their positions $\R_i$ and atomic numbers $Z_i$.
The forces driving their dynamics are derivatives of this energy, $\boldsymbol{F}_i = -\partial E / \partial \R_i$.
In tandem with the increasing availability of quantum-mechanical reference data, \mlips~\citep{bp07} have emerged as a data-driven approximation to this surface.
Modern \mlips are, almost without exception, \gnns~\citep{bhlp18pre} acting on a geometric graph of atoms whose edges encode interatomic vectors within a fixed cutoff radius $r_\text{c}$, with each atomic energy contribution $E_i$ local to a receptive field that grows linearly with the number of message-passing iterations. The total energy is a sum of $N$ such contributions, each with cost bounded by the finite cutoff, so one forward pass is $\bigo{N}$.\footnote{The relevant scaling for \mlips is $N \rightarrow \infty$ at constant density (the \emph{thermodynamic limit}), so the work per atomic environment stays constant.}
Recently, a line of \emph{foundation} \mlips has been trained on broad slices of the periodic table and is able to predict energies and forces for arbitrary chemistries in a zero-shot setting~\citep{bbzc25,mblc25,wduz25,yhhl24pre,rvdn25pre,mbcm26pre}.

\paragraph{Phonons}
Materials are modeled as periodic systems: only the atoms of the \emph{unit cell}, which is tiled in space, move independently; a \emph{supercell} combines several unit cells into a larger one to increase the number of independent degrees of freedom.
The \emph{interatomic force constants} $\Phi_{i\alpha,j\beta} = \partial^2 E / \partial \Rc_{i\alpha} \partial \Rc_{j\beta}$, the second-order expansion coefficients of the \pes around a minimum, are simply the Hessian of the energy with respect to Cartesian positions.
We write $\|\Phi_{ij}\|$ for the Frobenius norm of the $3\times3$ block of atom pair $(i,j)$.
Mass-weighting yields the dynamical matrix, whose eigendecomposition gives the squared phonon frequencies $\omega_s^2$.
The simulation cell must be large enough to contain all interactions between atoms; for \mlips, the required size can be determined exactly from the model's interaction range (\cref{sec:apx-data}).
Harmonic observables are thermal averages over the phonon spectrum~\citep{d93}; we consider the isochoric heat capacity%
\begin{equation}
\label{eq:cv}
\CV(T) = k_\mathrm{B} \sum_s \frac{x_s^2\, e^{x_s}}{(e^{x_s} - 1)^2}, \qquad x_s = \frac{\hbar \omega_s}{k_\mathrm{B} T}
\end{equation}
as an example, but other harmonic observables are readily available, such as Debye--Waller factors.

\section{Related work}
\label{sec:related}

\paragraph{Automatic differentiation for Hessians}
\citet{grm25} computed the Hessian of \mace with \ad, predicting zero-shot heat capacities on \citet{mnns22}'s porous-materials benchmark. Using dense \ad, they report a ceiling of ${\sim}1900$ atoms on an A100 GPU. We tackle this bottleneck.

\paragraph{Approximating Hessians}
Hessians can be used as a \emph{training} signal for \mlips: \citet{rsm25} showed that training on full \dft Hessians improves transition-state and vibrational-spectrum prediction, while PFT~\citep{kgms26}, PHL~\citep{rsbm26pre}, and the HORM dataset~\citep{chdz26} make this tractable via stochastic sampling of Hessian columns or Hutchinson-style random \hvps.

\paragraph{Predicting Hessians with \mlips}
HIP~\citep{btba25pre} bypasses \ad, directly predicting Hessians from SE(3)-equivariant features, with $10$--$70\times$ speedups over \ad on small molecules; the predicted Hessians are not guaranteed to be consistent with the gradient of the model's own forces.
Inconsistencies of this kind can lead to problems in physical simulations~\citep{blc25}.

\paragraph{Exploiting sparsity and decay}
Truncating force constants by distance is routine in lattice dynamics, whether through the supercell of a finite-displacement calculation~\citep{t23} or explicit cutoffs in force-constant fitting~\citep{es08,efe19}.
Probing methods in numerical linear algebra similarly reconstruct matrices with known sparsity or decay from few matrix-vector products~\citep{fss21,acmm26}.

\section{Methods}
\label{sec:methods}

\begin{figure}[t]
  \centering
  \includegraphics{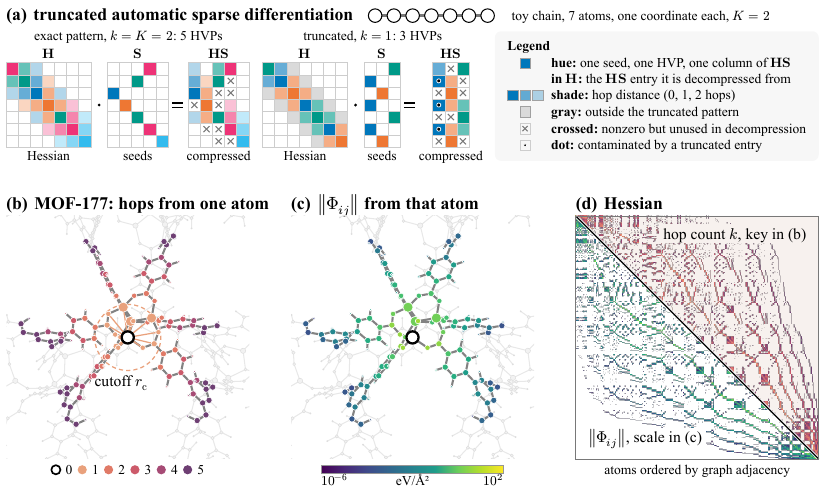}
  \caption{
  Truncated \ac{asd}.
  (a) Toy problem: a chain with $7$ atoms and a model with Hessian reach of $K=2$ hops. The Hessian $\hess$ is decompressed from \hvps with the columns of a seed matrix $\seed$, which yield the compressed product $\comp$. Since $\hess$ is sparse, star coloring yields $5$ colors (and thus $5$ \hvps), fewer than the $7$ of a dense evaluation.
  Every entry of $\hess$ is decompressed from one entry of $\comp$, shown by its hue, either directly or through its symmetric partner; entries of $\comp$ that are sums of several Hessian entries are never read (crossed).
  Truncating the sparsity pattern to one hop needs only three \hvps, but the neglected two-hop couplings (gray) now contaminate entries used for decompression (dots), leading to errors (\cref{sec:apx-errsplit}).
  (b) MOF-177 with \pet-XS: the atoms within $K = 5$ hops of one Zn atom on the model's input graph. We indicate the atom's cutoff sphere and one-hop edges. The rest of the cell is drawn in gray.
  (c) The same atoms colored by the force-constant block norm $\|\Phi_{ij}\|$ between each atom and the marked one, showing its decay with distance.
  (d) The whole Hessian of MOF-177 with one cell per pair of atoms, ordered so that graph neighbors are adjacent. Hop count above the diagonal, $\|\Phi_{ij}\|$ below, structural zeros white.
  }
  \label{fig:overview}
\end{figure}

\paragraph{Hessians via \ad}
Given a function implemented as a computer program, \ad automatically generates programs that evaluate \jvps or \vjps at the same asymptotic cost as the function itself \citep{gw08}.
For an \mlip energy $E: \reals^{3N} \rightarrow \reals$, one \vjp therefore yields the entire gradient, and with it all forces, at the cost of one energy evaluation (the \emph{cheap gradient principle}, \citealp{w82,bs83}).
Differentiating that program again yields an \hvp $\hess \V$, again at the cost of a small multiple of $E$ \citep{p94}.\footnote{We refer to \citet{davm24} for a detailed comparison of \hvp modes in \jax and \torch.}
Individual products are cheap, but materializing $\hess$ is not: each \hvp $\hess \mathbf{e}_i$ with a standard basis vector recovers one column, so the full Hessian costs $3N$ passes; at $\bigo{N}$ per pass for a linear-scaling \mlip, this amounts to $\bigo{N^2}$.
Phonon observables require the full eigenspectrum, and hence the materialized Hessian---matrix-free approaches cannot be used.
\paragraph{Automatic sparse differentiation}%
\asd via \emph{compressed evaluation}~\citep{cpr74,pt79}, recently re-popularized in the context of machine learning~\citep{hdm25,hd25}, reduces the $3N$ \hvp passes to the number $c$ of \emph{symmetrically orthogonal partitions} of the Hessian sparsity pattern, provided the pattern is known ahead of time.
Each pass evaluates one \hvp with the sum of a partition's basis vectors, a column of the seed matrix $\seed$, returning the sum of the corresponding columns, a column of $\comp$; orthogonality lets individual entries be read off that sum (\cref{fig:overview}a).
The partitioning is found by graph coloring, specifically \emph{star coloring}~\citep{cm84,gmp05,gtpw09}, so we call $c$ the number of \emph{colors}; for very sparse patterns, $c \ll 3N$.
The sparsity pattern need not be exact: a superset of the true one is safe and merely costs extra colors, while a subset breaks orthogonality and makes the recovery lossy. This tradeoff is the basis of the truncated \asd discussed below.
We use \texttt{asdex}~\citep{hd26} for coloring and decompression.
Coloring in this particular case can be sped up by observing that the sparsity pattern is made up of $3 \times 3$ blocks, one per pair of atoms, spanning the three components of each atom's position, with the relevant sparsity fully determined by atom-atom connectivity. We therefore only color the graph of atoms, and then assign the three coordinates of each atom three distinct colors derived from the atom's color. This yields a valid star coloring of the full sparsity pattern at a fraction of the cost (see \cref{sec:apx-coloring}).

\paragraph{Sparsity detection for \mlips}
Using \asd requires knowing the sparsity pattern of the target derivative matrix without evaluating it.
For \mlips based on \gnns, the sparsity pattern follows directly from the input graph.
Manual or empirical inspection of the model determines the receptive field of its energy terms: each term depends on the atoms within $L$ hops of the atom (\mace) or edge (\pet) it is predicted for, with $L$ the number of message-passing layers.
Since the second derivative $\partial^2 E / \partial \R_i \partial \R_j$ couples two atoms only when both lie in the receptive field of a shared energy term, the Hessian's sparsity pattern extends at most to the diameter of that field: $K = 2L$ hops for \mace{} and $K = 2L + 1$ for \pet; both counts are derived and verified in \cref{sec:apx-hops}.
The $K$-hop sparsity pattern is then given by the nonzero entries of
\begin{equation}
\label{eq:pattern}
\adj^{(K)} = \sum_{k=0}^{K} \adj^{k},
\end{equation}
evaluated in Boolean arithmetic, where $\adj$ is the adjacency matrix of the input graph, $\adj_{ij} = 1$ if atoms $i$ and $j$ share an edge (\cref{fig:overview}b--d).
This pattern is \emph{exact}: entries outside it are structurally zero.
It is also the graph the coloring acts on.

\paragraph{Truncated automatic sparse differentiation}

Truncating \cref{eq:pattern} at $k < K$ yields a nested family of increasingly sparse patterns that retain couplings only up to $k$ hops on the input graph.
Truncated patterns reduce the cost of \asd, since sparser patterns admit colorings with fewer colors, and therefore require fewer \hvps and less storage.
The resulting Hessian is approximate in two ways: couplings beyond $k$ hops are discarded, and, since star coloring guarantees collision-free recovery only for a conservative sparsity pattern, the neglected couplings contaminate retained entries sharing a color.
Both errors are controlled by the magnitude of the neglected couplings.
Because message-passing \mlips build up longer-ranged interactions by iterating local ones, these couplings should decay with graph distance, mirroring the decay of the underlying \pes.
Whether they decay fast enough for truncated Hessians to remain useful is an empirical question, answered in \cref{sec:experiments}.%

\section{Experiments}
\label{sec:experiments}

\paragraph{Setup}

We evaluate three foundation \mlips: \mace (MACE-MP-0 medium, $L{=}2$, $K{=}4$), \pet-XS ($L{=}2$, $K{=}5$), and \pet-S ($L{=}3$, $K{=}7$); details are given in \cref{sec:apx-models}.
We study porous materials (\acsp{mof}, \acsp{cof}, zeolites) from \citet{mnns22}'s benchmark: a stratified twelve-structure subset in supercells converged with respect to each model's interaction range (both described in \cref{sec:apx-data}); for validation against prior work, we recompute all 233 structures in their unit cells with \mace{} (\cref{sec:apx-cv-pipeline}).
We additionally consider giant \acs{mof} unit cells: MIL-101 (3604 atoms), one of the largest unit cells among common \acsp{mof}, as well as MIL-100 (2788 atoms), MOF-210 (1854 atoms), and MOF-177 (808 atoms).
All Hessians are computed in single precision (a choice ablated in \cref{sec:apx-cv-pipeline}) on a single H100 GPU; heat capacities follow the pipeline described there.

\paragraph{Hessian entries decay with graph distance}
We first examine the premise of truncation: \cref{fig:decay} shows the magnitude of the force-constant blocks $\|\Phi_{ij}\|$ as a function of the hops $k$ between atoms $i$ and $j$. Entries decay by roughly two orders of magnitude per hop for \mace{} and slightly more than one for the \pet{} models, so that at $k{=}3$ the median block norm lies $3.5$ to $7$ orders of magnitude below the on-site blocks. The speed of decay depends on chemistry: \acsp{mof} and \acsp{cof} decay at a similar rate, while zeolites display a slower decay.

\paragraph{Automatic sparse differentiation is exact}
\looseness=-1
Next, we use both dense and sparse \ad to compute the exact Hessians for two kinds of systems: the giant \acsp{mof}, for most of which no supercells are required to obtain converged vibrational properties, and smaller systems, which must be tiled up to each model's effective interaction range.
Cost (measured end-to-end) and accuracy for the giant \acsp{mof} are listed in \cref{tab:giants}; results for all structures in \cref{tab:cost}.
In all cases, \asd{} reproduces the dense predictions to within \num{5e-4} in relative Frobenius norm, and typically to \num{e-6}, close to expectation in single precision.
However, it offers modest speedups at best: median end-to-end speedups of $1.7\times$ for \mace{}, $1.4\times$ for \pet-XS and $1.2\times$ for \pet-S, never more than $6\times$ anywhere, and around break-even for the densest structures.
The reason for this is a lack of usable sparsity: the full interaction range of the \mlips{} used in this work is comparable to even the large \acs{mof} unit cells, and the supercells used for smaller systems are constructed to have exactly the minimum size to contain that interaction radius.
The only additional source of sparsity is therefore porosity, i.e., regions where the input graph is genuinely disconnected, which varies strongly between systems.
Low sparsity means many colors, so there is little to gain from \asd.
Predicted heat capacities are compared with published calorimetry in \cref{sec:apx-exp}, where all models overestimate the measured values.

\begin{table}[t]
  \caption{Dense, truncated ($k{<}K$), and exact ($k{=}K$, shaded) Hessians of the giant \acsp{mof}. $N$ is the number of atoms in the model-converged supercell. Speedups are relative to dense wall time, all end to end, including sparsity pattern construction and coloring. $\delta C_V$ is the deviation of $\CV$ at \SI{300}{\kelvin} from the dense reference in per mille, with the acoustic sum rule enforced. \mace{} has $K{=}4$.}
  \label{tab:giants}
  \centering
  \footnotesize
  \setlength{\tabcolsep}{3.5pt}
\begin{tabular}{llrr rrrr>{\columncolor{exactshade}}r rrrr>{\columncolor{exactshade}}r}
\toprule
& & & Time & \multicolumn{5}{c}{Speedup over dense} & \multicolumn{5}{c}{$\delta C_V$ (\textperthousand)} \\
\cmidrule(lr){5-9} \cmidrule(lr){10-14}
& & & dense & \multicolumn{4}{c}{truncated} & exact & \multicolumn{4}{c}{truncated} & exact \\
\cmidrule(lr){5-8} \cmidrule(lr){9-9} \cmidrule(lr){10-13} \cmidrule(lr){14-14}
       Structure  &             Model  &               $N$  &    (\si{\second})  &           $k{=}1$  &           $k{=}2$  &           $k{=}3$  &           $k{=}4$  &               $K$  &           $k{=}1$  &           $k{=}2$  &           $k{=}3$  &           $k{=}4$  &               $K$ \\ 
\midrule
         MIL-101  &     \textsc{Mace}  &        \num{3604}  &        \num{3529}  &        \num{36.5}  &        \num{13.3}  &         \num{5.9}  &                --  &         \num{2.7}  &      \num{-20.77}  &       \num{-0.18}  &       \num{<0.01}  &                --  &       \num{<0.01} \\ 
                  &   \textsc{Pet}-XS  &        \num{3604}  &         \num{316}  &         \num{8.9}  &         \num{8.5}  &         \num{7.3}  &         \num{6.3}  &         \num{5.3}  &     \num{-109.50}  &      \num{-15.09}  &       \num{<0.01}  &       \num{<0.01}  &       \num{<0.01} \\ 
                  &    \textsc{Pet}-S  &        \num{3604}  &        \num{1248}  &        \num{23.9}  &        \num{16.6}  &        \num{11.1}  &         \num{7.5}  &         \num{1.9}  &      \num{-55.58}  &       \num{-0.22}  &       \num{<0.01}  &       \num{<0.01}  &       \num{<0.01} \\ 
\midrule
         MIL-100  &     \textsc{Mace}  &        \num{2788}  &        \num{1246}  &        \num{18.6}  &         \num{7.5}  &         \num{3.1}  &                --  &         \num{1.3}  &       \num{-7.71}  &       \num{<0.01}  &       \num{<0.01}  &                --  &       \num{<0.01} \\ 
                  &   \textsc{Pet}-XS  &        \num{2788}  &         \num{227}  &         \num{6.3}  &         \num{5.9}  &         \num{5.2}  &         \num{4.4}  &         \num{3.6}  &      \num{-87.99}  &       \num{-5.60}  &       \num{<0.01}  &       \num{<0.01}  &       \num{<0.01} \\ 
                  &    \textsc{Pet}-S  &        \num{2788}  &         \num{876}  &        \num{16.8}  &        \num{11.5}  &         \num{7.7}  &         \num{5.0}  &         \num{1.2}  &      \num{-24.81}  &        \num{0.03}  &       \num{<0.01}  &       \num{<0.01}  &       \num{<0.01} \\ 
\midrule
         MOF-210  &     \textsc{Mace}  &        \num{1854}  &         \num{555}  &        \num{10.9}  &         \num{8.0}  &         \num{5.1}  &                --  &         \num{2.6}  &      \num{-32.40}  &       \num{-0.39}  &       \num{<0.01}  &                --  &       \num{<0.01} \\ 
                  &   \textsc{Pet}-XS  &        \num{1854}  &         \num{111}  &         \num{3.3}  &         \num{3.2}  &         \num{3.0}  &         \num{3.0}  &         \num{2.7}  &     \num{-127.99}  &      \num{-51.02}  &       \num{-8.55}  &       \num{<0.01}  &       \num{<0.01} \\ 
                  &    \textsc{Pet}-S  &        \num{1854}  &         \num{379}  &         \num{8.0}  &         \num{6.6}  &         \num{5.4}  &         \num{4.3}  &         \num{1.5}  &      \num{-76.44}  &      \num{-14.13}  &       \num{<0.01}  &       \num{<0.01}  &       \num{<0.01} \\ 
\midrule
         MOF-177  &     \textsc{Mace}  &        \num{6464}  &        \num{9431}  &        \num{80.5}  &        \num{30.2}  &        \num{14.3}  &                --  &         \num{5.6}  &      \num{-26.28}  &       \num{<0.01}  &       \num{<0.01}  &                --  &       \num{<0.01} \\ 
                  &   \textsc{Pet}-XS  &         \num{808}  &          \num{48}  &         \num{1.5}  &         \num{1.4}  &         \num{1.4}  &         \num{1.4}  &         \num{1.3}  &     \num{-126.79}  &      \num{-39.20}  &       \num{-7.21}  &       \num{<0.01}  &       \num{<0.01} \\ 
                  &    \textsc{Pet}-S  &        \num{6464}  &        \num{3900}  &        \num{67.1}  &        \num{46.4}  &        \num{23.4}  &        \num{15.4}  &         \num{3.2}  &      \num{-66.16}  &       \num{-7.37}  &       \num{<0.01}  &       \num{<0.01}  &       \num{<0.01} \\ 
\bottomrule
\end{tabular}
\end{table}

\begin{figure}[t]
  \centering
  \includegraphics{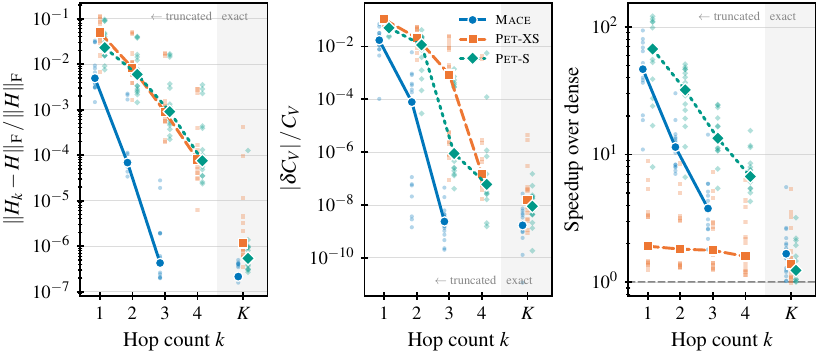}
  \caption{Truncation across the benchmark subset and the giant \acsp{mof}: relative Hessian error (left), relative $\CV$ error at \SI{300}{\kelvin} with the acoustic sum rule enforced (center), and end-to-end speedup over dense \ad (right), versus hop count $k$; each model's exact $K$ is set apart (shaded). Bold markers are medians over structures; small dots are individual structures. Sparse timings include pattern construction and coloring; the dashed gray line marks break-even.}
  \label{fig:truncation}
\end{figure}

\paragraph{Truncated automatic sparse differentiation trades accuracy for speed}
Finally, we measure Hessian error, $\CV$ error, and speedup of truncated sparsity patterns against the dense reference, at every hop count $k$ across the benchmark subset and the giant \acsp{mof} (\cref{fig:truncation}).
As the sparsity pattern becomes more and more filled in with increasing $k$, errors and speedups decrease.
Derived observables converge abruptly: at unconverged rungs the relative error in $\CV$ is comparable to that of the Hessian, and within two hops it drops to more than two orders of magnitude below the Hessian error.
With \SI{0.1}{\percent} as the threshold for $\CV$, \mace is converged at $k=2$ for the vast majority of structures, \pet-S at $k=3$, and \pet-XS at $k=4$, with corresponding median speedups of $11\times$, $13\times$, and $1.6\times$.
The exceptions are zeolites, as expected from the slower decay of the Hessian seen in \cref{fig:decay}. The slowest-converging zeolite requires one more hop: $k=3$ for \mace at $5.9\times$ speedup, $k=4$ for \pet-S at $5.4\times$, and the exact $k=K=5$ for \pet-XS, where little speedup remains.
The modest speedups for \pet-XS throughout reflect its high overall speed and small converged supercells---timings are dominated by fixed overheads and not \hvps.
The residual errors in converged $\CV$ are a threshold effect: the pipeline drops near-zero modes (\cref{sec:apx-cv-pipeline}), and truncation can move one across the threshold, adding or removing its ${\sim}k_\mathrm{B}$; at exact $K$, $\CV$ agrees with the dense reference to better than \num{1e-5} relative.
The Hessian error itself separates into the two contributions identified in \cref{sec:methods}; discarded couplings dominate until the pattern is nearly converged (\cref{sec:apx-errsplit}).

\section{Conclusion}
\label{sec:conclusion}

This work showed how to apply \asd to compute the Hessians of \mlips. We explained how to obtain the sparsity pattern of \mpnn-based \mlips in closed form from their input graph by summing powers of the adjacency matrix. 
We then applied \asd to porous materials, whose large unit cells and porosity make them a natural best case for sparse Hessians. However, the interaction ranges of current foundation models are comparable to even these large unit cells, so \asd reproduces dense Hessians faithfully but at only modest speedups.

This motivated a truncated version of \asd{}.
The exact sparsity pattern is the endpoint of a hierarchy of sparser patterns that only keep interactions between atoms separated by fewer graph hops.
Using these sparser patterns neglects terms of the Hessian, but allows fewer colors, and hence fewer \hvps, for its materialization.
We find that this approximation yields similar observables at a fraction of the computational cost.
This somewhat surprising result is due to the decay of the Hessian with distance: truncating interactions yields a controlled loss of accuracy, which can be reduced by increasing the number of hops to be considered.

Several directions follow from here.
Higher-order derivatives are the natural next step: three-phonon linewidths require contractions of the third-derivative tensor with a fixed direction, which reverse-mode \ad evaluates without forming the tensor, at a small multiple of the Hessian's cost~\citep{gg16}. This tensor inherits the sparsity of the Hessian, so sparsity patterns, colorings~\citep{dn19}, and truncations of this work carry over directly; how it is best computed with higher-order and Taylor-mode techniques~\citep{gw08,guw00,bjd19} is left open.
Truncation will not be as benign there as for heat capacity: linewidths weight precisely the distant couplings that truncation discards, so the sufficient hop count will depend on the observable.
A theoretical analysis of the truncation error, connecting it to the decay rates of \cref{fig:decay}, would put the choice of hop count on firmer ground for all derivative orders.

Our results also expose message-passing depth as a performance-relevant design decision. The receptive field, so far chosen with accuracy in mind and little cost attached, directly sets the sparsity of higher-order derivatives, and keeping it compact now has a large payoff.
For models whose interaction range is formally infinite, whether through deep receptive fields or explicit long-range terms, the Hessian is fully dense.
Nevertheless, as physical interactions decay with distance, truncated \asd can still lead to large speedups where exact \asd cannot.
Finally, since the predictive pipelines shown here are differentiable, at least in principle, they also pave the way toward fine-tuning on \emph{experimental} instead of simulation data.

The presented truncated (or lossy) approach to \asd potentially also extends beyond the domain of \mlips. For any derivative operator whose per-entry magnitude can be estimated ahead of time, truncated \asd materializes it at reduced cost by neglecting small, but nonzero, entries.

\section*{Code and data availability}
All code and data of this work are available in a single archive at \url{https://doi.org/10.5281/zenodo.22813524}, mirrored at \url{https://github.com/sirmarcel/tasd4mlip-archive}.
The archive contains the \jax{} re-implementation of \mace, the experiment scripts, and the relaxed structures, run records, and derived observables of every experiment.
A production-ready implementation of truncated \asd will be made available in \texttt{pet-jax}~\citep{ls26}.
Further information and an overview of data sources can be found at \url{https://marcel.science/tasd4mlip}.

\begin{ack}
MFL acknowledges funding from the German Research Foundation (DFG), project number 544947822.
MFL and MC acknowledge funding from the European Research Council (ERC) under the European Union's Horizon 2020 research and innovation programme (Grant Agreement 101001890-FIAMMA) and from NCCR Separations, a National Centre of Competence in Research funded by the Swiss National Science Foundation (grant number 229280).
Furthermore, AH gratefully acknowledges funding from the German Federal Ministry of Education and Research under the grant BIFOLD26B.
\end{ack}

\bibliographystyle{plainnat}
\bibliography{babel}

\newpage
\appendix
\crefalias{section}{appendix}
\crefalias{subsection}{subappendix}

\section{Benchmark structures}
\label{sec:apx-data}

\paragraph{Benchmark set}

All heat-capacity benchmarks draw on the 233-structure porous-materials set assembled by \citet{mnns22}, comprising 214 \acsp{mof}, 9 \acsp{cof}, and 9 zeolites (one further structure carries no class label), for which \citet{grm25} provide reference \mace heat capacities and phonon frequencies in unit cells and $2{\times}2{\times}2$ supercells.
We take the published crystal structures as inputs and relax them within our own pipeline (\cref{sec:apx-cv-pipeline}); no geometries are inherited from either upstream source.%

\paragraph{Subset selection}

Computing converged Hessians for all 233 structures would be wasteful, so we benchmark on subsets, chosen deterministically as follows.
Within each class, structures are ordered by farthest-point sampling~\citep{ks69}: starting from the most central structure, the next pick is always the one most different from all previous ones.
Distance is measured on three rank-transformed features: the logarithm of the unit-cell atom count, the number density, and the Hessian fill in the minimal exact supercell, which stand in for size, porosity, and the sparsity available to our method.
The per-class orders are interleaved at 8 \acsp{mof} : 1 \acs{cof} : 1 zeolite per ten ranks, so that subsets span all classes even though the full set is dominated by \acsp{mof}.
A subset of size $n$ is the first $n$ entries of this ranking.
The benchmarked twelve extend the first ten ranks (the \emph{ten-structure subset} of \cref{sec:apx-cv-pipeline}): RSM1885 is skipped as infeasible in its \mace-converged supercell, and one further \acs{cof} and two zeolites are added for class coverage.%

\paragraph{Supercell selection}

Vibrational properties converge only once the simulation cell contains every interaction of the model, so supercells are chosen per structure and model.
We build the model's one-hop graph through its production input pipeline and propagate it to the Hessian hop count $K$ (\cref{sec:methods}), tracking the periodic image that each interaction reaches.
A supercell is admissible if no two of these interactions fold onto the same force-constant entry, which reduces to a divisibility test on image-offset differences.
Among admissible cells, we pick the diagonal multiplier yielding the smallest volume; the search is exhaustive but finite, since the multiplier one larger than the largest offset difference in each direction is always admissible.
Common practice instead interpolates the dynamical matrix in reciprocal space from a supercell judged large enough; we use exact cells to keep the pipeline simple and directly comparable to \citet{grm25}.
Both the graph and $K$ depend on the model, so the converged supercell, and with it $N$ in \cref{tab:cost}, does too.%

\paragraph{Giant \acsp{mof}}

For benchmarks beyond typical unit-cell sizes we use four \acsp{mof} with large primitive cells: MOF-177 (808 atoms), MOF-210 (1854), MIL-100 (2788), and MIL-101 (3604), taken unmodified from the structure library of RASPA2~\citep{dces16} and relaxed in our pipeline like all other structures.
For MOF-177 and MIL-101, experimental heat capacities from calorimetry are available for comparison (\cref{sec:apx-exp}).%

\section{Model implementations}
\label{sec:apx-models}

We use two model families in this work: MACE-MP-0 medium~\citep{bkoc22,bbzc25} (\mace) and PET-MAD-1.5~\citep{pc23,bpsc26,mbcm26pre} in its XS and S variants (\pet-XS, \pet-S).

\mace{} combines equivariant message passing with higher-order Atomic Cluster Expansion features~\citep{bkoc22}.
The MP-0 medium foundation model has $L{=}2$ message-passing layers at a \SI{6}{\angstrom} cutoff, giving $K{=}4$, and is trained on Materials Project relaxation trajectories computed at the PBE(+$U$) level of \dft~\citep{bbzc25}.

\pet{} is an edge-to-edge transformer (within local neighborhoods) without built-in rotational equivariance~\citep{pc23}.
The PET-MAD-1.5 models~\citep{mbcm26pre} are fine-tuned from the PET-OMat models of \citet{bpsc26}, which are pretrained on the OMat24 dataset~\citep{bszu26}.
The fine-tuning set is the smaller, curated MAD-1.5 dataset~\citep{mbcm26pre}, computed at the \rscan level of \dft in a non-magnetic setting.
The XS and S variants have $L{=}2$ and $L{=}3$ layers, giving $K{=}5$ and $K{=}7$.
Within a fixed outer cutoff radius, \pet{} shrinks its effective cutoff per atomic environment to bound the neighbor count.
Since that cutoff depends on atomic positions, it contributes additional force terms, which we disregard: they couple atoms beyond the selected neighbor list and would decrease the very sparsity this work depends on (see the ablations in \cref{sec:apx-cv-pipeline}).

Both models are distributed for \torch; to compute \hvps efficiently with \jax's \ad system, we rely on \jax{} re-implementations, using \texttt{pet-jax}~\citep{ls26} for \pet{} and implementing \mace{} ourselves.
We verify these on the first 50 structures of our benchmark ranking (\cref{sec:apx-data}), comparing energies, forces and stresses in single and double precision against the upstream \torch{} models in double precision (\cref{tab:parity}).
In double precision the re-implementations agree with upstream to far below the accuracy of the models themselves; in single precision, both stacks depart from that reference by the same amount, so the implementation contributes no error beyond rounding.
One structure deviates further because a pair distance happens to fall where the two backends' \texttt{tanh} kernels saturate differently, a property of the floating-point implementation rather than of either model.
Allowing \jax{} to use TF32 for matrix multiplications (the default behavior on GPUs) is a different matter: it leads to errors nearly four orders of magnitude higher in the forces compared to full single precision, and is therefore not used in this work.

\begin{table}[h]
  \caption{Deviation of each prediction path from the upstream \torch{} model in double precision, over the first 50 structures of the benchmark ranking (\cref{sec:apx-data}).
Rows per model: upstream in single precision, then our \jax{} re-implementation in double precision, in full single precision (the production setting of this work), and with TF32 matrix multiplications.
Cells are the median over the 50 structures with the worst structure in parentheses, energies per atom, forces and stresses as the largest deviating component.
The first three rows are measured on CPU, the TF32 row on GPU. Running the full single-precision path on the GPU instead yields the same deviations to within rounding noise, so the last row isolates TF32, not the device.}
  \label{tab:parity}
  \centering
  \footnotesize
  \setlength{\tabcolsep}{3.5pt}
\begin{tabular}{ll rrr}
\toprule
                             Model  &                          Prediction  &                $\Delta E$ (eV/atom)  &                 $\Delta F$ (eV/\AA)  &        $\Delta \sigma$ (eV/\AA$^3$) \\ 
\midrule
                     \textsc{Mace}  &                       \torch{} fp32  &       \num{3.3e-06} (\num{5.4e-06})  &       \num{4.0e-05} (\num{8.4e-05})  &       \num{5.5e-08} (\num{2.0e-07}) \\ 
                                    &                         \jax{} fp64  &       \num{1.2e-08} (\num{3.0e-08})  &       \num{3.6e-07} (\num{6.1e-07})  &       \num{5.9e-09} (\num{1.7e-08}) \\ 
                                    &                         \jax{} fp32  &       \num{1.0e-07} (\num{3.7e-07})  &       \num{4.0e-05} (\num{8.4e-05})  &       \num{5.5e-08} (\num{1.8e-07}) \\ 
                                    &                   \jax{} fp32, TF32  &       \num{8.3e-05} (\num{5.0e-04})  &       \num{2.9e-01} (\num{6.4e-01})  &       \num{4.5e-04} (\num{1.1e-03}) \\ 
\midrule
                   \textsc{Pet}-XS  &                       \torch{} fp32  &       \num{2.1e-06} (\num{7.1e-06})  &       \num{4.1e-05} (\num{9.0e-05})  &       \num{6.4e-08} (\num{2.2e-07}) \\ 
                                    &                         \jax{} fp64  &       \num{9.8e-07} (\num{2.1e-06})  &       \num{8.0e-06} (\num{2.3e-05})  &       \num{6.4e-08} (\num{2.0e-07}) \\ 
                                    &                         \jax{} fp32  &       \num{1.1e-06} (\num{2.3e-06})  &       \num{4.4e-05} (\num{5.6e-04})  &       \num{1.2e-07} (\num{7.1e-07}) \\ 
                                    &                   \jax{} fp32, TF32  &       \num{4.8e-04} (\num{1.0e-03})  &       \num{3.2e-01} (\num{6.5e-01})  &       \num{4.6e-04} (\num{1.2e-03}) \\ 
\midrule
                    \textsc{Pet}-S  &                       \torch{} fp32  &       \num{2.0e-06} (\num{7.2e-06})  &       \num{4.3e-05} (\num{8.8e-05})  &       \num{5.7e-08} (\num{1.9e-07}) \\ 
                                    &                         \jax{} fp64  &       \num{2.4e-07} (\num{4.6e-07})  &       \num{2.6e-06} (\num{4.8e-06})  &       \num{2.5e-08} (\num{8.0e-08}) \\ 
                                    &                         \jax{} fp32  &       \num{8.1e-08} (\num{2.8e-07})  &       \num{4.4e-05} (\num{8.6e-05})  &       \num{7.5e-08} (\num{2.3e-07}) \\ 
                                    &                   \jax{} fp32, TF32  &       \num{1.5e-04} (\num{6.0e-04})  &       \num{3.2e-01} (\num{6.7e-01})  &       \num{4.6e-04} (\num{1.2e-03}) \\ 
\bottomrule
\end{tabular}
\end{table}

\section{Verifying the Hessian hop count}
\label{sec:apx-hops}

The two counts follow from the readout style: a node-readout energy term depends on the $L$-hop ball around its atom, of graph diameter $2L$, while an edge-readout term depends on the union of the $L$-hop balls around two adjacent atoms, of diameter $2L+1$.
We confirm them by measuring double-precision Hessians of small hand-built structures (chains, rings, chains with second-neighbor bonds, chains with pendant leaves) whose connectivity we control exactly and whose graph diameter exceeds the predicted $K$, taking the measured $K$ as the largest graph distance at which any $3\times3$ Hessian block is nonzero.
Freshly initialized models, whose random weights make every structurally allowed coupling generically nonzero, match the predicted count in every one of 96 measurements spanning $L = 1,\dots,4$, both readout styles, all four graph families, and three seeds; the production checkpoints reproduce it at their trained depths, giving $K=4$ for MACE-MP-0 medium, $K=5$ for \pet-XS, and $K=7$ for \pet-S.
Beyond $K$, blocks are exactly zero.
For \pet, this is the hop count \emph{after} the adaptive cutoff procedure, whose position dependence our sparse pipeline discards (ablated in \cref{sec:apx-cv-pipeline}).

\section{Coloring on the atom graph}
\label{sec:apx-coloring}

Two coordinates are coupled in the Hessian pattern if and only if their atoms are within $K$ hops on the input graph.
The coordinate sparsity pattern is therefore the $N \times N$ atom pattern with every nonzero replaced by a dense $3 \times 3$ block: sparsity is fully determined by pairs of atoms, and coordinates only add a block structure.
For this reason, it suffices to color the atom sparsity pattern, which has one ninth of the nonzeros, and then give coordinate $d \in \{0, 1, 2\}$ of atom $i$ the color $3 c_i + d$, where $c_i$ is the atom's color.
By construction, this is a valid star coloring of the full coordinate pattern.

The timings we report combine two runs.
Hessians and their \hvps come from a run that colored the coordinate pattern directly, the coloring time from a later run that colored the atom graph only.
On all 234 sparse rungs of the benchmark the two colorings are identical, yielding the same \hvps, and only the cost of finding the coloring changes.
At each model's exact $K$, coloring the atom graph is $3$ to $22\times$ faster than coloring the coordinate pattern, $17\times$ in the median, and over the whole benchmark the coloring stage drops from \SI{22.8}{\hour} to \SI{1.2}{\hour}.

\section{Computing heat capacity}
\label{sec:apx-cv-pipeline}

\paragraph{Pipeline}
To compute heat capacity we follow the example of \citet{grm25}: initial unit cells are relaxed with BFGS over a \texttt{FrechetCellFilter} (L-BFGS with line search for the giant \acsp{mof}) to a maximum force of \SI[per-mode=symbol]{5e-3}{\electronvolt\per\angstrom} in double precision, with \mace using the D3 correction and \pet as-is. The Hessian is mass-weighted and diagonalized to give the frequencies entering \cref{eq:cv}, zeroing imaginary frequencies and dropping near-zero modes with wavenumber $|\nu| < \SI{e-3}{\per\centi\meter}$. In comparisons between sparse and dense Hessians (\cref{tab:cost,fig:truncation}) we additionally enforce the acoustic sum rule, so that the acoustic modes cancel exactly instead of falling on either side of the drop threshold. We verify the pipeline by recomputing frequencies and heat capacities for all 233 structures in their unit cells with \mace{}, comparing against the published values of \citet{grm25}.
The two pipelines differ in the relaxed geometries and in the D3 Hessian, which is a finite difference over forces there and computed with \ad here.
The deviations are small: the largest frequency deviation within a structure is \SI{0.7}{\per\centi\meter} in the median over structures, and $\CV$ at \SI{300}{\kelvin} deviates by \num{3e-5} relative in the median, with a worst case of $2.6\%$.
Cells beyond $2{\times}2{\times}2$ were out of computational reach for \citet{grm25}.

\paragraph{Ablating precision, the D3 correction, and adaptive-cutoff forces}
Our pipeline departs from the reference Hessian calculations of \citet{grm25} in three ways: \mlips are evaluated in single rather than double precision, \mace{} Hessians omit the D3 dispersion correction, and \pet{} Hessians disregard the force contributions of the adaptive cutoff procedure (\cref{sec:apx-models}).
To measure what each departure costs, we compute dense unit-cell Hessians for the ten-structure subset (\cref{sec:apx-data}), flipping one axis at a time away from the reference configuration (\mace{} in double precision with D3, \pet{} in double precision with adaptive-cutoff forces), and compare frequencies, Hessians, and heat capacities with the acoustic sum rule enforced.
The single-precision ablation is run twice, with matrix multiplications in full single precision and in the TF32 matmul mode that \jax{} defaults to on GPUs (\cref{sec:apx-models}).
\Cref{tab:ablations} summarizes the results.
Single precision itself is harmless: with full matrix multiplications, heat capacities deviate by less than $\ablFpMedCeil\,\%$ in the median.
TF32 is not, and its effect is of the same size as omitting D3 or the adaptive-cutoff forces.
All production Hessians in this work therefore pin matrix multiplications to full single precision, which makes each \hvp 15--29\,\% slower for \mace{} and 55--97\,\% slower for \pet{} than with TF32, at production supercell sizes on an H100.
The production configuration, combining all three departures, deviates from the reference by at most $\ablProdMedHi\,\%$ in the median, with worst cases of a few percent.
These worst cases are a threshold effect, not a shift of the spectrum: some structures have genuine low-lying optical modes near the drop threshold of the pipeline, and a perturbation of any origin can push one across it, adding or removing that mode's ${\sim}k_\mathrm{B}$ from $\CV$, a few percent in a small unit cell.
The acoustic sum rule removes this ambiguity for the acoustic modes only.
Every deviation above $1\,\%$ in the ablation is of this kind, with a single exception: for \ablDispStruct, omitting D3 genuinely shifts the spectrum, by up to \SI{\ablDispShift}{\per\centi\meter}, for a $\ablDispShiftCv\,\%$ change in $\CV$.

\begin{table}[t]
  \caption{Ablations of the production Hessian pipeline against the reference configuration of \citet{grm25}, computed as dense unit-cell Hessians for the ten-structure subset (\cref{sec:apx-data}).
Per ablation and model: median and worst relative deviation of $\CV$ at \SI{300}{\kelvin} with the acoustic sum rule enforced, worst single-mode frequency deviation, and median relative Frobenius error of the Hessian itself, each over the ten structures.
The first two rows split single precision by matrix multiplication mode, full fp32 (the production setting) and TF32. The last row combines all departures and is the production configuration.}
  \label{tab:ablations}
  \centering
  \footnotesize
  \setlength{\tabcolsep}{3.5pt}
\begin{tabular}{ll rrrr}
\toprule
        Ablation  &             Model  &  med.\ $|\delta C_V|$ (\%)  &  max $|\delta C_V|$ (\%)  &  max $|\Delta\nu|$ (cm$^{-1}$)  &  med.\ $\lVert\Delta\hess\rVert_F / \lVert\hess\rVert_F$ \\ 
\midrule
fp32 (full matmuls)  &     \textsc{Mace}  &     \num{6.9e-06}  &     \num{3.6e-05}  &       \num{0.019}  &     \num{1.8e-06} \\ 
                  &   \textsc{Pet}-XS  &     \num{4.7e-05}  &     \num{2.4e-02}  &         \num{4.1}  &     \num{2.6e-06} \\ 
                  &    \textsc{Pet}-S  &     \num{3.3e-05}  &     \num{9.1e-04}  &        \num{0.14}  &     \num{2.1e-06} \\ 
\midrule
fp32 (TF32 matmuls)  &     \textsc{Mace}  &       \num{0.010}  &         \num{3.0}  &          \num{51}  &     \num{5.3e-03} \\ 
                  &   \textsc{Pet}-XS  &       \num{0.029}  &         \num{3.0}  &         \num{112}  &     \num{4.9e-03} \\ 
                  &    \textsc{Pet}-S  &       \num{0.015}  &         \num{1.6}  &          \num{26}  &     \num{3.4e-03} \\ 
\midrule
           no D3  &     \textsc{Mace}  &       \num{0.056}  &         \num{2.9}  &          \num{67}  &     \num{3.6e-03} \\ 
\midrule
no adaptive-cutoff forces  &   \textsc{Pet}-XS  &        \num{0.12}  &         \num{2.5}  &         \num{377}  &     \num{7.0e-03} \\ 
                  &    \textsc{Pet}-S  &       \num{0.035}  &         \num{1.4}  &          \num{23}  &     \num{1.9e-03} \\ 
\midrule
            fp32  &     \textsc{Mace}  &       \num{0.056}  &         \num{2.9}  &          \num{67}  &     \num{3.6e-03} \\ 
         + no D3  &   \textsc{Pet}-XS  &        \num{0.12}  &         \num{2.5}  &         \num{374}  &     \num{7.0e-03} \\ 
+ no adaptive-cutoff forces  &    \textsc{Pet}-S  &       \num{0.035}  &         \num{1.4}  &          \num{23}  &     \num{1.9e-03} \\ 
\bottomrule
\end{tabular}
\end{table}

\clearpage
\section{Cost and accuracy across the benchmark}
\label{sec:apx-table}

\Cref{tab:cost} lists wall times and heat-capacity errors at every hop count for every structure--model pair in the benchmark.

\begin{table}[H]
  \caption{Wall time and accuracy of dense, exact ($K$), and truncated ($k{<}K$) Hessians for every benchmark structure and model.
$N$ is the number of atoms in the supercell, chosen per model as the smallest multiple of the unit cell that accommodates its full interaction range (\cref{sec:apx-data}); rows are grouped by chemistry class and sorted by the largest supercell across models within each group.
Sparse wall times are end to end, including sparsity pattern construction and coloring as well as the Hessian evaluation itself.
$\delta\CV$ is the relative deviation of the heat capacity at \SI{300}{\kelvin} from the same pair's dense reference, computed as in \cref{sec:apx-cv-pipeline} with the acoustic sum rule enforced; $0.00$ marks magnitudes below $0.005\%$.
\mace{} has $K{=}4$, so its $k{=}4$ entries appear in the $K$ column.}
  \label{tab:cost}
  \centering
  \footnotesize
  \setlength{\tabcolsep}{2.3pt}
  \setlength{\defaultaddspace}{2.2pt}
  \renewcommand{\arraystretch}{0.85}
\begin{tabular}{lllr rrrrrr rrrrr}
\toprule
& & & & \multicolumn{6}{c}{Wall time (\si{\second})} & \multicolumn{5}{c}{$\delta C_V$ (\%)} \\
\cmidrule(lr){5-10} \cmidrule(lr){11-15}
           Class  &         Structure  &             Model  &               $N$  &             dense  &           $k{=}1$  &           $k{=}2$  &           $k{=}3$  &           $k{=}4$  &               $K$  &           $k{=}1$  &           $k{=}2$  &           $k{=}3$  &           $k{=}4$  &               $K$ \\ 
\midrule
             MOF  &           MOF-210  &     \textsc{Mace}  &        \num{1854}  &         \num{555}  &          \num{51}  &          \num{69}  &         \num{109}  &                --  &         \num{213}  &       \num{-3.24}  &       \num{-0.04}  &        \num{0.00}  &                --  &        \num{0.00} \\ 
                  &                    &   \textsc{Pet}-XS  &        \num{1854}  &         \num{111}  &          \num{34}  &          \num{35}  &          \num{37}  &          \num{37}  &          \num{42}  &      \num{-12.80}  &       \num{-5.10}  &       \num{-0.85}  &        \num{0.00}  &        \num{0.00} \\ 
                  &                    &    \textsc{Pet}-S  &        \num{1854}  &         \num{379}  &          \num{48}  &          \num{57}  &          \num{70}  &          \num{88}  &         \num{260}  &       \num{-7.64}  &       \num{-1.41}  &        \num{0.00}  &        \num{0.00}  &        \num{0.00} \\ 
\addlinespace
                  &           MIL-100  &     \textsc{Mace}  &        \num{2788}  &        \num{1246}  &          \num{67}  &         \num{167}  &         \num{402}  &                --  &         \num{968}  &       \num{-0.77}  &        \num{0.00}  &        \num{0.00}  &                --  &        \num{0.00} \\ 
                  &                    &   \textsc{Pet}-XS  &        \num{2788}  &         \num{227}  &          \num{36}  &          \num{38}  &          \num{44}  &          \num{52}  &          \num{63}  &       \num{-8.80}  &       \num{-0.56}  &        \num{0.00}  &        \num{0.00}  &        \num{0.00} \\ 
                  &                    &    \textsc{Pet}-S  &        \num{2788}  &         \num{876}  &          \num{52}  &          \num{76}  &         \num{113}  &         \num{175}  &         \num{701}  &       \num{-2.48}  &        \num{0.00}  &        \num{0.00}  &        \num{0.00}  &        \num{0.00} \\ 
\addlinespace
                  &           MIL-101  &     \textsc{Mace}  &        \num{3604}  &        \num{3529}  &          \num{97}  &         \num{265}  &         \num{597}  &                --  &        \num{1327}  &       \num{-2.08}  &       \num{-0.02}  &        \num{0.00}  &                --  &        \num{0.00} \\ 
                  &                    &   \textsc{Pet}-XS  &        \num{3604}  &         \num{316}  &          \num{35}  &          \num{37}  &          \num{43}  &          \num{50}  &          \num{59}  &      \num{-10.95}  &       \num{-1.51}  &        \num{0.00}  &        \num{0.00}  &        \num{0.00} \\ 
                  &                    &    \textsc{Pet}-S  &        \num{3604}  &        \num{1248}  &          \num{52}  &          \num{75}  &         \num{112}  &         \num{166}  &         \num{643}  &       \num{-5.56}  &       \num{-0.02}  &        \num{0.00}  &        \num{0.00}  &        \num{0.00} \\ 
\addlinespace
                  &           RSM1876  &     \textsc{Mace}  &        \num{3969}  &        \num{4291}  &         \num{107}  &         \num{345}  &         \num{990}  &                --  &        \num{2788}  &       \num{-1.70}  &        \num{0.00}  &        \num{0.00}  &                --  &        \num{0.00} \\ 
                  &                    &   \textsc{Pet}-XS  &        \num{1176}  &          \num{67}  &          \num{34}  &          \num{36}  &          \num{37}  &          \num{40}  &          \num{43}  &       \num{-8.93}  &       \num{-1.21}  &        \num{0.00}  &        \num{0.00}  &        \num{0.00} \\ 
                  &                    &    \textsc{Pet}-S  &        \num{3969}  &        \num{1500}  &          \num{54}  &          \num{78}  &         \num{140}  &         \num{247}  &        \num{1328}  &       \num{-5.07}  &       \num{-1.19}  &        \num{0.00}  &        \num{0.00}  &        \num{0.00} \\ 
\addlinespace
                  &           RSM0023  &     \textsc{Mace}  &        \num{3840}  &        \num{4624}  &         \num{121}  &         \num{557}  &        \num{1769}  &                --  &        \num{4160}  &       \num{-1.19}  &       \num{-0.03}  &        \num{0.00}  &                --  &        \num{0.00} \\ 
                  &                    &   \textsc{Pet}-XS  &         \num{768}  &          \num{45}  &          \num{33}  &          \num{35}  &          \num{35}  &          \num{39}  &          \num{45}  &       \num{-5.94}  &       \num{-0.86}  &       \num{-0.19}  &        \num{0.00}  &        \num{0.00} \\ 
                  &                    &    \textsc{Pet}-S  &        \num{4800}  &        \num{2202}  &          \num{56}  &          \num{99}  &         \num{206}  &         \num{408}  &        \num{2198}  &       \num{-3.73}  &       \num{-0.06}  &       \num{-0.03}  &        \num{0.00}  &        \num{0.00} \\ 
\addlinespace
                  &           MOF-177  &     \textsc{Mace}  &        \num{6464}  &        \num{9431}  &         \num{117}  &         \num{312}  &         \num{659}  &                --  &        \num{1698}  &       \num{-2.63}  &        \num{0.00}  &        \num{0.00}  &                --  &        \num{0.00} \\ 
                  &                    &   \textsc{Pet}-XS  &         \num{808}  &          \num{48}  &          \num{32}  &          \num{34}  &          \num{34}  &          \num{35}  &          \num{37}  &      \num{-12.68}  &       \num{-3.92}  &       \num{-0.72}  &        \num{0.00}  &        \num{0.00} \\ 
                  &                    &    \textsc{Pet}-S  &        \num{6464}  &        \num{3900}  &          \num{58}  &          \num{84}  &         \num{167}  &         \num{253}  &        \num{1220}  &       \num{-6.62}  &       \num{-0.74}  &        \num{0.00}  &        \num{0.00}  &        \num{0.00} \\ 
\addlinespace
                  &           RSM1877  &     \textsc{Mace}  &        \num{6615}  &       \num{23285}  &         \num{490}  &        \num{2788}  &       \num{10680}  &                --  &       \num{22759}  &       \num{-1.15}  &       \num{-0.02}  &        \num{0.00}  &                --  &        \num{0.00} \\ 
                  &                    &   \textsc{Pet}-XS  &        \num{1764}  &         \num{116}  &          \num{35}  &          \num{38}  &          \num{42}  &          \num{53}  &          \num{75}  &      \num{-13.80}  &       \num{-2.78}  &        \num{0.00}  &        \num{0.00}  &        \num{0.00} \\ 
                  &                    &    \textsc{Pet}-S  &        \num{6615}  &        \num{4699}  &          \num{65}  &         \num{145}  &         \num{340}  &         \num{742}  &        \num{4458}  &       \num{-4.98}  &       \num{-1.10}  &        \num{0.00}  &        \num{0.00}  &        \num{0.00} \\ 
\addlinespace
                  &           RSM1162  &     \textsc{Mace}  &        \num{2916}  &        \num{1137}  &          \num{56}  &         \num{143}  &         \num{439}  &                --  &         \num{948}  &       \num{-2.40}  &       \num{-0.08}  &        \num{0.00}  &                --  &        \num{0.00} \\ 
                  &                    &   \textsc{Pet}-XS  &         \num{864}  &          \num{45}  &          \num{31}  &          \num{33}  &          \num{33}  &          \num{37}  &          \num{40}  &      \num{-13.51}  &       \num{-1.54}  &        \num{0.00}  &        \num{0.00}  &        \num{0.00} \\ 
                  &                    &    \textsc{Pet}-S  &        \num{6912}  &        \num{4439}  &          \num{62}  &         \num{110}  &         \num{230}  &         \num{519}  &        \num{2782}  &       \num{-4.44}  &       \num{-2.13}  &        \num{0.00}  &        \num{0.00}  &        \num{0.00} \\ 
\addlinespace
                  &           RSM0010  &     \textsc{Mace}  &        \num{7700}  &       \num{39619}  &         \num{839}  &        \num{5622}  &       \num{22015}  &                --  &       \num{38950}  &       \num{-0.01}  &       \num{-0.01}  &        \num{0.00}  &                --  &        \num{0.00} \\ 
                  &                    &   \textsc{Pet}-XS  &        \num{2112}  &         \num{113}  &          \num{35}  &          \num{37}  &          \num{43}  &          \num{56}  &          \num{78}  &       \num{-8.39}  &       \num{-1.26}  &        \num{0.00}  &        \num{0.00}  &        \num{0.00} \\ 
                  &                    &    \textsc{Pet}-S  &        \num{8448}  &        \num{6845}  &          \num{69}  &         \num{187}  &         \num{481}  &        \num{1031}  &        \num{6552}  &       \num{-7.05}  &       \num{-0.39}  &       \num{-0.05}  &        \num{0.00}  &        \num{0.00} \\ 
\addlinespace
                  &           RSM0274  &     \textsc{Mace}  &        \num{7840}  &       \num{33275}  &         \num{515}  &        \num{3220}  &       \num{11851}  &                --  &       \num{26961}  &       \num{-0.45}  &        \num{0.00}  &        \num{0.00}  &                --  &        \num{0.00} \\ 
                  &                    &   \textsc{Pet}-XS  &        \num{2016}  &         \num{124}  &          \num{35}  &          \num{38}  &          \num{43}  &          \num{55}  &          \num{71}  &       \num{-7.79}  &       \num{-0.31}  &       \num{-0.17}  &        \num{0.00}  &        \num{0.00} \\ 
                  &                    &    \textsc{Pet}-S  &        \num{8960}  &        \num{8723}  &          \num{76}  &         \num{187}  &         \num{522}  &        \num{1094}  &        \num{7283}  &       \num{-4.04}  &       \num{-0.13}  &       \num{-0.04}  &        \num{0.00}  &        \num{0.00} \\ 
\addlinespace
                  &           RSM0047  &     \textsc{Mace}  &        \num{3888}  &        \num{3022}  &          \num{68}  &         \num{209}  &         \num{636}  &                --  &        \num{1503}  &       \num{-1.76}  &        \num{0.00}  &        \num{0.00}  &                --  &        \num{0.00} \\ 
                  &                    &   \textsc{Pet}-XS  &        \num{1152}  &          \num{64}  &          \num{34}  &          \num{36}  &          \num{39}  &          \num{44}  &          \num{53}  &       \num{-5.38}  &       \num{-1.30}  &        \num{0.00}  &        \num{0.00}  &        \num{0.00} \\ 
                  &                    &    \textsc{Pet}-S  &        \num{9216}  &        \num{6967}  &          \num{63}  &         \num{136}  &         \num{283}  &         \num{593}  &        \num{3291}  &       \num{-7.45}  &       \num{-2.15}  &       \num{-0.03}  &        \num{0.00}  &        \num{0.00} \\ 
\midrule
             COF  &           20561N3  &     \textsc{Mace}  &        \num{7680}  &       \num{21529}  &         \num{285}  &        \num{1678}  &        \num{5410}  &                --  &       \num{11001}  &       \num{-0.79}  &        \num{0.00}  &        \num{0.00}  &                --  &        \num{0.00} \\ 
                  &                    &   \textsc{Pet}-XS  &         \num{960}  &          \num{48}  &          \num{34}  &          \num{34}  &          \num{34}  &          \num{38}  &          \num{45}  &      \num{-10.07}  &       \num{-2.71}  &        \num{0.00}  &        \num{0.00}  &        \num{0.00} \\ 
                  &                    &    \textsc{Pet}-S  &        \num{5760}  &        \num{3091}  &          \num{59}  &         \num{107}  &         \num{242}  &         \num{454}  &        \num{2372}  &       \num{-5.51}  &       \num{-0.78}  &        \num{0.00}  &        \num{0.00}  &        \num{0.00} \\ 
\addlinespace
                  &           18150N2  &     \textsc{Mace}  &        \num{5850}  &       \num{13406}  &         \num{257}  &        \num{1168}  &        \num{3666}  &                --  &        \num{8754}  &       \num{-1.10}  &        \num{0.00}  &        \num{0.00}  &                --  &        \num{0.00} \\ 
                  &                    &   \textsc{Pet}-XS  &        \num{3120}  &         \num{200}  &          \num{37}  &          \num{40}  &          \num{44}  &          \num{51}  &          \num{61}  &      \num{-11.14}  &       \num{-3.53}  &       \num{-1.04}  &        \num{0.00}  &        \num{0.00} \\ 
                  &                    &    \textsc{Pet}-S  &        \num{9360}  &        \num{9426}  &          \num{78}  &         \num{195}  &         \num{388}  &         \num{827}  &        \num{4280}  &       \num{-3.90}  &       \num{-2.41}  &        \num{0.00}  &        \num{0.00}  &        \num{0.00} \\ 
\midrule
         zeolite  &               VFI  &     \textsc{Mace}  &        \num{4860}  &        \num{7759}  &         \num{168}  &         \num{681}  &        \num{2116}  &                --  &        \num{4212}  &       \num{-3.20}  &       \num{-0.09}  &        \num{0.00}  &                --  &        \num{0.00} \\ 
                  &                    &   \textsc{Pet}-XS  &        \num{1296}  &          \num{64}  &          \num{34}  &          \num{36}  &          \num{37}  &          \num{42}  &          \num{49}  &      \num{-12.31}  &       \num{-4.67}  &       \num{-1.55}  &       \num{-0.33}  &        \num{0.00} \\ 
                  &                    &    \textsc{Pet}-S  &        \num{5832}  &        \num{3657}  &          \num{66}  &         \num{147}  &         \num{324}  &         \num{671}  &        \num{3616}  &       \num{-2.31}  &       \num{-1.91}  &       \num{-0.14}  &        \num{0.00}  &        \num{0.00} \\ 
\addlinespace
                  &               NPT  &     \textsc{Mace}  &        \num{6912}  &       \num{11453}  &         \num{122}  &         \num{614}  &        \num{1946}  &                --  &        \num{5182}  &       \num{-7.03}  &       \num{-0.60}  &        \num{0.00}  &                --  &        \num{0.00} \\ 
                  &                    &   \textsc{Pet}-XS  &         \num{864}  &          \num{45}  &          \num{33}  &          \num{34}  &          \num{36}  &          \num{38}  &          \num{41}  &      \num{-10.15}  &       \num{-5.25}  &       \num{-2.36}  &        \num{0.00}  &        \num{0.00} \\ 
                  &                    &    \textsc{Pet}-S  &        \num{6912}  &        \num{4428}  &          \num{66}  &         \num{139}  &         \num{340}  &         \num{687}  &        \num{3656}  &       \num{-5.07}  &       \num{-4.18}  &        \num{0.04}  &       \num{-0.56}  &        \num{0.00} \\ 
\addlinespace
                  &               AFI  &     \textsc{Mace}  &        \num{6912}  &       \num{16600}  &         \num{240}  &        \num{1214}  &        \num{4274}  &                --  &        \num{9249}  &       \num{-5.60}  &       \num{-0.01}  &        \num{0.00}  &                --  &        \num{0.00} \\ 
                  &                    &   \textsc{Pet}-XS  &         \num{864}  &          \num{44}  &          \num{36}  &          \num{34}  &          \num{35}  &          \num{39}  &          \num{43}  &      \num{-12.04}  &       \num{-5.63}  &       \num{-2.48}  &       \num{-0.42}  &        \num{0.00} \\ 
                  &                    &    \textsc{Pet}-S  &        \num{9216}  &        \num{8079}  &          \num{74}  &         \num{186}  &         \num{489}  &        \num{1063}  &        \num{6618}  &       \num{-6.10}  &       \num{-2.38}  &       \num{-0.07}  &       \num{-0.01}  &        \num{0.00} \\ 
\bottomrule
\end{tabular}
\end{table}

\section{Force-constant decay}
\label{sec:apx-decay}

\begin{figure}[h]
  \centering
  \includegraphics{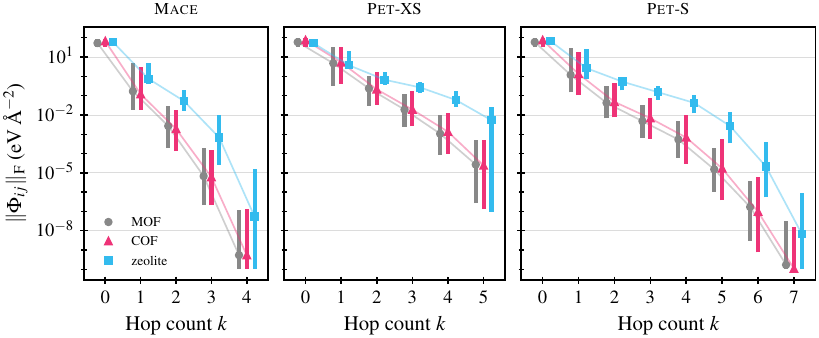}
  \caption{Force-constant block norms $\|\Phi_{ij}\|$ versus hop count $k$, split by chemistry class, one panel per model: median and 10--90\% band over all coupled pairs of the benchmark subset and the giant \acsp{mof}, plus RSM1885 for the \pet{} models.}
  \label{fig:decay}
\end{figure}

\Cref{fig:decay} is computed from the dense Hessians of the benchmark subset and the giant \acsp{mof} in their model-converged supercells (\cref{tab:cost}), plus RSM1885 for the \pet{} models; its \mace-converged supercell is infeasible (\cref{sec:apx-data}).
Only rows belonging to atoms in the original unit cell appear, as replica atoms are equivalent and carry the same force constants.
The hop count of a pair $(i,j)$ is the smallest $k$ for which $(\adj^k)_{ij}$ is nonzero, with $\adj$ the input-graph adjacency of \cref{sec:methods}; $k{=}0$ labels the on-site blocks, and $\lVert\Phi_{ij}\rVert$ is the Frobenius norm of the pair's $3\times3$ force-constant block.
Quantiles are taken over the pooled pairs of each chemistry class, with the giant \acsp{mof} counted as \acsp{mof}.
The same Hessians confirm the sparsity pattern at production scale: beyond each model's $K$, every block is exactly zero.

\clearpage
\section{Truncation error decomposition}
\label{sec:apx-errsplit}

\begin{figure}[H]
  \centering
  \includegraphics{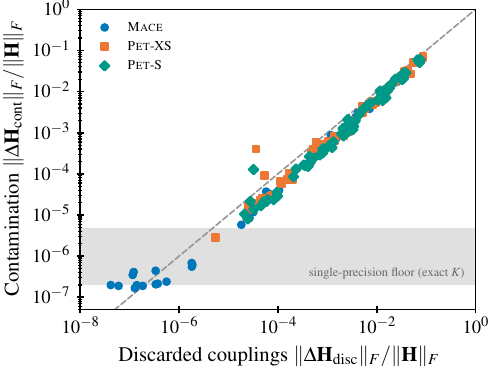}
  \caption{Contamination of the retained entries against the discarded couplings, both relative to the dense Hessian norm: one marker per (structure, model, hop count) for every truncated pattern of \cref{tab:cost} and of RSM1885 for the \pet{} models. The dashed line is equality; the band spans the central 80\% of the exact-$K$ errors, the single-precision noise floor.}
  \label{fig:errsplit}
\end{figure}

A truncated Hessian is not the dense Hessian with distant couplings set to zero: star coloring prevents collisions only within the assumed sparsity pattern, so neglected couplings contribute to retained entries (\cref{sec:methods}).
The error therefore has two parts: the \emph{discarded} couplings themselves, and the \emph{contamination} of the retained entries.
No entry appears in both, so their squared Frobenius norms add up to the squared total error.
We compute the split from the stored Hessians of every sparse pattern in \cref{tab:cost}, plus RSM1885 for the \pet{} models as in \cref{sec:apx-decay}; the Hessians are single precision, while the comparison is done in double precision.
\Cref{fig:errsplit} plots one against the other: contamination tracks the discarded couplings at a roughly constant fraction, a median of $0.60$ across four orders of magnitude, until it bottoms out at the single-precision noise floor measured by the exact-$K$ patterns, where floating-point noise, not truncation, sets the error.

\clearpage
\section{Comparison with experiment}
\label{sec:apx-exp}

\begin{figure}[H]
  \centering
  \includegraphics{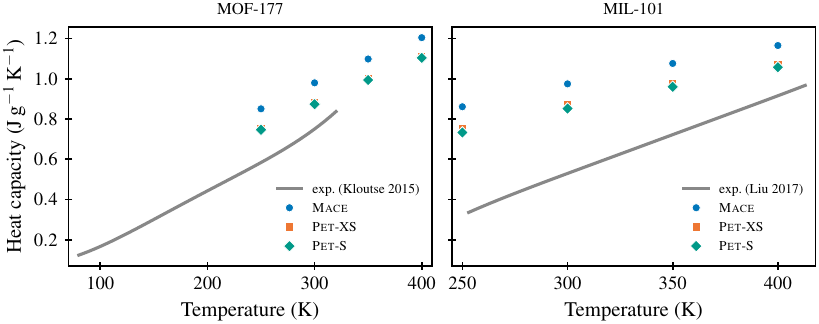}
  \caption{Predicted heat capacity against calorimetry for the two giant \acsp{mof}: measured $C_p(T)$ curves (gray; MOF-177 from \citealp{kzcc15}, MIL-101 from \citealp{lxzz17}) and $\CV$ from dense \mlip{} Hessians in converged supercells (markers), with the acoustic sum rule enforced.}
  \label{fig:calorimetry}
\end{figure}

For the two giant \acsp{mof} with published calorimetric heat capacities~\citep{kzcc15,lxzz17}, \cref{fig:calorimetry} compares predictions from dense Hessians in converged supercells against the measured curves; since $C_p \approx \CV$ for these solids in this temperature range, the comparison is direct.
All models overestimate: at \SI{300}{\kelvin}, \mace{} lies $30\%$ above experiment for MOF-177 and the \pet{} models around $17\%$, consistent with the systematic over-softening of predicted phonon spectra observed by \citet{grm25}.
For MIL-101 the gap widens to $84\%$ and $61$--$64\%$, too large to attribute to the harmonic approximation or the potentials alone; the measured sample, with defects and possible residual guest species, likely differs substantially from the ideal crystal we compute.
Discrepancies of this kind are precisely what large-scale Hessian calculations make visible, and what fine-tuning on experimental data would correct.

\end{document}